# Institutional Platform for Secure Self-Service Large Language Model Exploration

**V. K. Cody Bumgardner, PhD[1], Mitchell A. Klusty, BS[1], W. Vaiden Logan, BS[1], Samuel E. Armstrong, MS[1], Caroline N. Leach, BS[1], Kenneth L. Calvert, PhD[1], Caylin Hickey, BS[1], Jeff Talbert, PhD[1]**

**[1]University of Kentucky, Lexington, KY, USA**

**Abstract**

*This paper introduces a user-friendly platform developed by the University of Kentucky Center for Applied AI, designed to make customized large language models (LLMs) more accessible. By capitalizing on recent advancements in multi-LoRA inference, the system efficiently accommodates custom adapters for a diverse range of users and projects. The paper outlines the system's architecture and key features, encompassing dataset curation, model training, secure inference, and text-based feature extraction. We illustrate the establishment of a tenant-aware computational network using agent-based methods, securely utilizing islands of isolated resources as a unified system. The platform strives to deliver secure, affordable LLM services, emphasizing process and data isolation, end-to-end encryption, and role-based resource authentication. This contribution aligns with the overarching goal of enabling simplified access to cutting-edge AI models and technology in support of scientific discovery and the development of biomedical informatics.*

## Introduction

Generative pre-trained transformers (GPT) have garnered great research and public interest. Driven by the public availability and ground-breaking performance of ChatGPT,[1] an incredibly broad set of use cases and research areas are being developed. Generative models, such as Stable Diffusion[2] and multi-modal models like GPT-4[3] further expand the generative reach of AI through audio, images, and videos. While the hosted infrastructure of ChatGPT/GPT-4 provides broad accessibility, users' ability to adapt the service to their own domain was initially limited to how they instructed (input content and structure) the model to respond, a process commonly referred to as "prompt engineering"[4]. Shortly after the release of ChatGPT, Meta publicly released Llama[5] and later Llama 2 and Llama 3,[6] which, unlike ChatGPT and other commercial services offerings, could not only be used locally but also trained for specific domains and fine-tuned for specific problems. While the practice of publicly releasing model architectures and pre-trained weights was already common with BERT[7] and other derivative efforts, the size of the Llama foundational model and the associated computational cost of training such large language models (LLMs) were unprecedented. At the time of writing, there are dozens of popular foundational LLMs and tens of thousands of derivative models[8]. Unfortunately, generative tasks are difficult to evaluate, and benchmarks can be easily manipulated[9]. Therefore, model training processes and associated dataset generation require a careful understanding of the problem domain and often human evaluation.

In large part, the success of the LLM ecosystem is due to the coordination of open-source libraries, models, and tools by the Hugging Face[10] community. The Hugging Face Transformers[11] library has become the de facto standard for LLM implementations, allowing for broad interoperability and rapid adoption of new models and associated technologies. Numerous transformers-based platforms and tools, such as Large Model Systems Organization's (LMSYS) FastChat[12] and OpenAccess AI Collective's Axolotl[13] have been developed for the in-house training and evaluation of LLMs. Likewise, cloud-based services such as AWS SageMaker[14] and Microsoft's Machine Learning Services[15] implement standard Hugging Face libraries and interoperable model standards. In addition to standard libraries, projects such as vLLM[16] provide multi-user inference services with OpenAI-compatible API interfaces. Support for the OpenAI API allows frameworks that are powered by language models, like LangChain,[17] to seamlessly make use of commercial and local LLMs. Most recently, Microsoft announced the Azure OpenAI Service,[18] allowing users to better adapt OpenAI models to their data and tasks. Through the combination of commercial and open-source efforts, a wide range of options are available for LLM training and inference. However, in either case, users are responsible for the curation, integration, and protection of their data and associated models, which can influence what resources are available to specific use cases.

At the time of writing, commercial offerings like OpenAI's GPT-4 dominate evaluation leaderboards, such as the LMsys Chatbot Arena.[19] Despite the generally superior performance of proprietary models, a large and active open-source LLM community has emerged. Along with standard supervised fine-tuning,[20] numerous LLM technologies and techniques have developed in attempts to meet or exceed the capabilities and performance of commercial services.

For example, additional pre-training[21] with unstructured data and instruction fine-tuning with structured data can be used to add domain-specific information to existing foundational models. In addition, Reinforcement Learning with Human Feedback (RLHF) techniques such as Reward Modeling,[22] Proximal Policy Optimization (PPO),[23] and Direct Preference Optimization (DPO)[24] are used to refine model behavior based on human preference. In addition to training techniques, dozens of innovative technologies have been implemented to both improve efficiency and expand the capabilities of available foundational models. For example, RoPE scaling[25] extends the maximum context length of pre-trained models, FlashAttention[26] reduces training resource requirements, and Low Rank Adaptation (LoRA) adapters[27] efficiently decouple user-trained parameters from the underlying larger base models. In addition, numerous quantization techniques such as GPTQ,[28] GGML,[29] qLoRA,[30] and AWQ[31] have been developed to reduce model weight resolution, and associated resource costs, while attempting to maintain model accuracy. The rate of local LLM community development, combined with a diverse set of techniques, tools, and development options, has created a learning curve that continues to grow steeper by the day, impeding the application of local LLMs to potentially valuable use cases.

Perhaps the greatest impediment to the broad operational deployment of local LLMs is the associated resource costs. A modest-sized 7 billion parameter model requires 14GB of GPU vRAM for inference when loaded at half resolution (fp16). As previously mentioned, model weights can be further quantized, reducing memory requirements. However, the operational cost of running hundreds or thousands of unique models is financially prohibitive for many use cases. As with previous non-LLM natural language processing models, tools such as NVIDIA's TensorRT-LLM[32] have been developed to optimize model inference performance, reduce computational cost, and manage parallelism for production models. Despite inference optimizations, resource costs associated with hosting large collections are significant. While LoRA techniques are used to decouple user-trained parameters from base models, the common practice is to merge LoRA and base weights, effectively creating a new, unique model. While only a fraction of the weights might have changed through LoRA training, once merged, the costs of the changed and unchanged weights are encumbered. This process and associated cost are repeated for each merged model, regardless of the global ratio of trained to duplicated parameters. Through a process called Segmented Gather Matrix-Vector multiplication (SGMV), a team of researchers was able to demonstrate the efficient use of LoRA adapters without the need to first merge training and existing weights, as demonstrated in the Punica[33] project. Derivative works such as S-LoRA[34] have demonstrated the ability to host thousands of independent adapters concurrently from the same base model. This is accomplished using a unified memory pool to dynamically manage adapter weights with differing ranks and varying sequence lengths. The efficiencies gained through multi-LoRA efforts have reduced the resource costs of hosting large collections of models by orders of magnitude. Sharing the cost of hosting a base model between adapters allows larger base models to be used while reducing the overall costs and increasing accessibility.

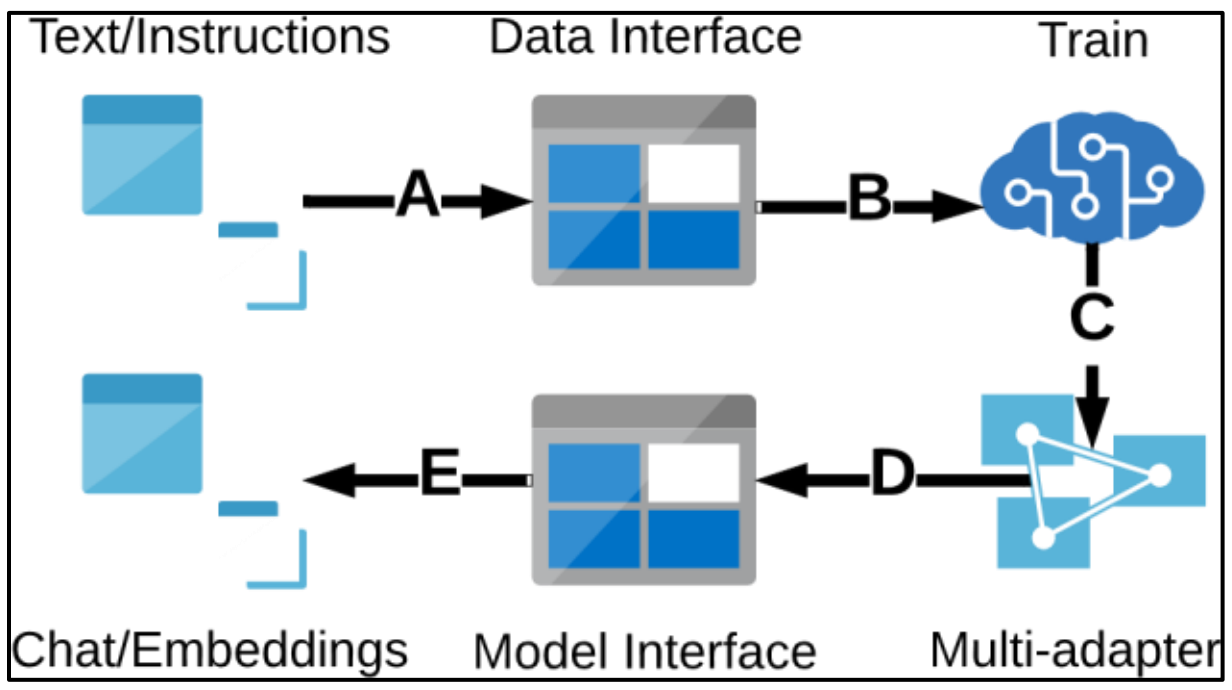

***Figure 1***. *High-level system overview of models*

Locally trained LLMs are well-suited for biomedical informatic applications that require stringent security and data access controls. Utilizing Protected Health Information (PHI) requires a HIPAA-compliant system that ensures patient privacy. This cannot be guaranteed by general publicly available commercial LLMs. Local training and hosting of base models and associated LoRA adapters allows organizations to securely use PHI data in training datasets and model queries, as those organizations can maintain the data management policies of the local models and ensure HIPAA standards are followed within their system. Training models on customized datasets also provides an opportunity to create a wide range of informatics tools with specialized purposes. Potential applications of such a model could be a trainer for drug counselors to simulate conversations with a patient, a tool for generating potential differential diagnoses using patient information, or even a tool that generates treatment recommendations based on the latest information. Local LLMs allow us to create models that work from private data and operate within sensitive areas that commercial models have been explicitly trained to avoid.

In a rapidly evolving area like generative AI, it can be difficult to identify the appropriate technologies for specific use cases and even harder to put those technologies into use. In this paper, we present a self-service platform for the secure end-to-end selection, training, evaluation, and private hosting of user-trained LLMs. Our system provides interactive dataset curation interfaces, model configuration, composition tools, and agent-based methods for the secure hosting of custom LLMs and associated adapters. This system operates within a secured environment that can be validated for HIPAA and more stringent compliance where necessary. It is designed to provide sustainably affordable access to private LLMs, embedding extraction/search services, and transcription models, ensuring users can train and deploy custom LLMs without the typical management complexity and overhead cost of hosting these models. The system was also designed with full OpenAI API compatibility, meaning it meets the industry standards of accessibility to AI tools and can easily be integrated with a wide range of tools and applications that were designed to work with OpenAI's models. This opens our system to the broader ecosystem of AI technologies.

## Methods

In this section, we will describe the components of LLM Factory, explain architectural considerations, and discuss key features. These features include exposing LLM inference through an OpenAI compatible API, training of LoRA adapters, multi-adapter inference, text-based feature extraction, and system security.

Figure 1 illustrates a high-level overview of our system's components and data flow. Figure labels correspond to the subsections presented in this section. Figure 2 shows a more detailed diagram of the various components of the system that permit the creation of LoRA adapters and querying of available models.

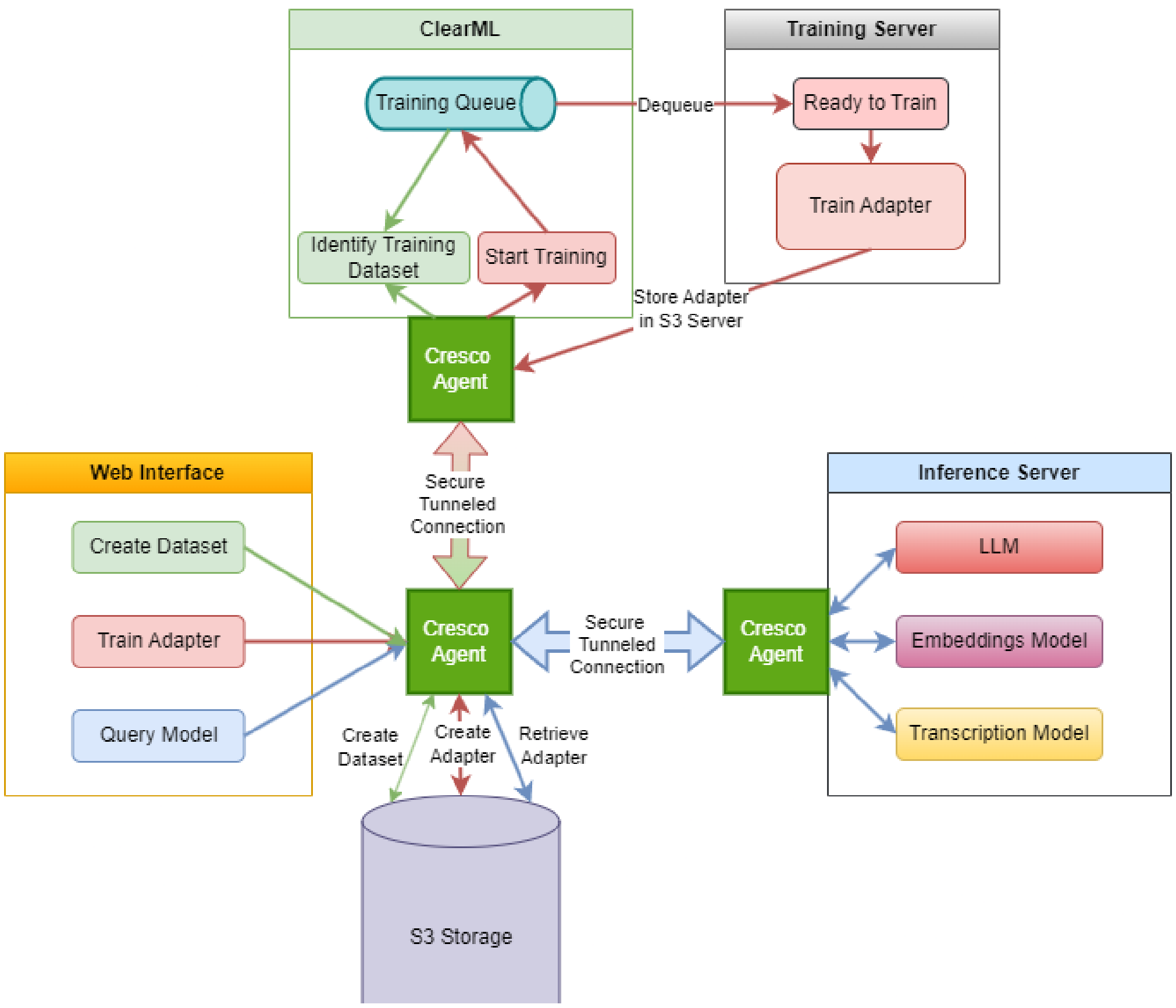

***Figure 2*** *System Overview Diagram*

### *Model Inference*

An LLM inference interface that is capable of switching between private models at time of execution is essential to the utility of this system. We began by creating an internal web interface that stores histories of chat sessions and gives the user the ability to adjust the system prompt, maximum output tokens, temperature, and the weights of selected adapters. It also can be put in "Comparison Mode," which shows side-by-side comparisons of the model output with and without the active adapters. This permits the user to test various configurations of their generated adapters and parameters of the model in an internal playground environment and store the sessions for later review.

More important than an internal interface is an externally accessible API. We make use of the LoRAX[35] server for model inference. LoRAX supports multi-LoRA inference through a custom REST API, Python client library, and OpenAI-compatible API. The multi-LoRA inference capabilities of LoRAX allow us to serve hundreds of private models in isolation from the same GPU, providing considerable efficiencies over standalone model deployments. Our OpenAI-compatible API allows our system to be used with other open and commercial libraries and applications that are OpenAI-compatible. We ensure security on this API by performing input format verification and content standard adherence before passing the request to LoRAX for inference. Security is further enhanced through an API key management system, which is used to verify user access to specific models and adapters. It also logs requests for our audit system to prevent abuse and track token usage. This system permits users to easily generate and delete API keys, allowing independence between each of the user's keys and swift rotation of compromised keys.

This API also permits users to create frontend interfaces that can be customized to their individual needs and utilize LLM Factory as a backend. For instance, we developed a WordPress plugin that allows seamless integration between LLM Factory and an interface deployable on WordPress sites. This capability provides users the flexibility to design unique user experiences tailored for specific tasks with the concerns of managing their own AI infrastructure alleviated by LLM Factory integration.

***Adapter Training***

LLM Factory enables efficient training of custom LoRA adapters, supporting both pretraining on unstructured text and generation of full adapters trained on structured JSON datasets. This system utilizes ClearML[35], a job management service to orchestrate training across standalone server training, local clusters, or cloud-based resources. In addition, ClearML provides hyperparameter tuning optimization and enforces a consistent training and evaluation environment by tracking source code, libraries, and data used in the training process.

Resulting LoRA adapters, related artifacts, and training datasets are stored in a secure, locally hosted S3-compatable server. This S3 server integrates with ClearML and LoRAX to ensure all data is only accessible where necessary and only transmitted when necessary, during training or inference. Data is never stored on or passed through ClearML, rather, there is a reference to the relevant S3 object locations that are passed to and then accessed by the training machine. LoRAX provides the capability to load adapters from an S3 source file extremely quickly, permitting the adapter files to only exist in the S3 server.

Our approach facilitates flexible and scalable development of LoRA adapters while ensuring compliance with security protocols, simplifying the process of creating fine-tuned LLMs, especially in research environments that benefit from tailored machine learning solutions.

We also expose a number of adapters to all users. These include several code generation adapters, which have been trained on producing reliable code from natural language inputs, a Neo4j Cypher query adapter, and an uncensored adapter, which allows the user to bypass restrictions on Llama 3 that prevent it from discussing sensitive topics. These adapters are widely applicable to many areas of research, so they have been shared with all users to expedite usage of LLM Factory.

***Multi-Adapter Inference***

At the time of writing, the dominant paradigm for LLM distribution and inference is to use monolithic models. While LoRA techniques are widely used to train model adapters efficiently, the resulting adapters are typically merged with base models prior to use. The result is that while custom training might only impact a small percentage of parameters, the full cost of model storage and computation is incurred for each independent instance.

As previously mentioned, in late 2023, researchers representing Punica and S-LoRA projects demonstrated the ability to host thousands of independent adapters concurrently from the same base model. This breakthrough in efficiency has enabled the cost-effective hosting of custom models in a way not previously possible, opening the door to much broader involvement in LLM efforts. The ability to support multi-adapter inference serves as the backbone of LLM Factory, allowing end-users to train and interact with their models directly with limited costs compared to previous techniques where the cost of each adapter matches the cost of hosting another base model.

We make use of the LoRAX[36] server for model inference. LoRAX supports multi-LoRA inference through a custom REST API, Python client library, and OpenAI-compatible API. Each LoRAX server instance is bound to a supported base model, such as Bloom, GPT2, Llama, Mistral, Mixtral, Phi, Qwen, and Zephyr.[37] In addition to being trained from a supported base model, adapters must be trained with LoRAX-supported targets,[38] the components of the model that will be affected by the adapter. A known list of inference servers with associated base models and candidate adapters is actively maintained by the system through a process that will be described under *Secure*

*Computational Network*. Incoming requests are routed to appropriate inference servers based on the requested adapter and associated base model. The prompt and additional configuration options are included at the time of execution, as shown in Figure 3.

```
{
  "inputs": "[INST] Generate SOAP notes from the following transcription... [/INST]",
  "parameters": {
        "temperature": 0.5,
        "max_new_tokens": 512,
        ...
        "adapter_id": "adapters/soap-node-generator",
        "adapter_source": "local"
  }
}
```

***Figure 3*** *Multi-LoRA Request*

API interactions between the user and the inference server are proxied through the system, where we isolate the users' traffic and inject the appropriate LoRA adapter based on the user's session and associated permissions. The requests are made through our secure tunneling system detailed in *Secure Computational Network*. This process holds true for both interactive chat and API-based access. The inference server request is regenerated for each request, allowing for the translation of API protocols and future migration between inference servers. Outside of Punica and S-LoRA conceptual demonstration, LoRAX was one of the first projects to support multi-LoRA. However, many popular inference servers, such as vLLM, are implementing multi-LoRA support. Likewise, as new adapter technologies, such as LoHa,[39] LoKr,[40] and LoftQ[41] are added to the supporting PEFT libraries, they are implemented in inference servers. This allows our system to offer cutting edge adapter support and rapidly switch between adapters and merge multiple together in ways that would not be possible with other inference servers.

***Vector Embeddings***

In addition to various chat completion models, LLM Factory provides access vector embeddings models. At the time of writing, LLM and embedding services are independent. Vector embeddings are numerical representations of objects, words, sentences, or concepts in a vector space which captures semantic relationships between the objects the vectors represent. Large sets of data/documents can be encoded into vectors and stored in a vector database[42]. A method called Retrieval Augmented Generation (RAG)[43] is used to search a vector database by encoding a search query and performing a Euclidean distance calculation to find the closest related documents in the vector database based on semantic similarities between the query and the documents in the database. The extracted entities can then be used as context for an LLM, meaning vector databases can serve as vast knowledge stores for language models, allowing them to generate informed, contextually relevant responses.

Embeddings provide a mechanism for enriching responses by injecting domain specific knowledge on a per-query basis. Unlike LoRA adapters which adjust the LLM itself to specialize it in a particular domain, embeddings using RAG techniques allow dynamic retrieval of relevant information, enhancing the model's knowledge without affecting its root language capabilities. This also provides security benefits for the encoded data. A custom permission schema can be designed to limit the scope of the search to only include specific vertexes from the database based on the access rights verified in the request. This allows for strict control over which users and applications can extract which records from the database.

In our system, we exposed an embeddings model through an API that give the means to vectorize datasets and queries, and, by extension, the means to perform similarity searches. We also wrote a Python library to integrate our system with the popular language model toolkit, LangChain. This library meshes the embeddings model of our system with LangChain to ensure consistent vectorization.

***Audio Transcriptions***

LLM Factory also hosts the OpenAI Whisper[44] transcription model, accessible through an OpenAI-compatible API endpoint. Whisper is an automatic speech recognition model that transcribes audio files into text. The model offers multilingual support and is noise-robust, meaning it can handle noisy environments. These qualities make the model extremely versatile for practical and research purposes. LLM Factory allows for fast transcription of short audio files (~30-60s) and provides the means to open a data stream to asynchronously transcribe longer audio files, or even transcribe live-recorded audio. Applications of transcriptions in this domain might include facilitating data collection, dictation of medical documentation, and transcription of clinical consultations. There are also applications for telemedicine conferences, such as automatic translation between patients and practitioners who speak different languages.

***Projects***

When working on a research project with LLM Factory, we envision that multiple team members may individually create adapters that should be accessible to the entire team. This necessitates a way to share adapters within the system. Our solution is a project management system. This allows users to create projects, assign roles within the project, associate adapters with the project that can be queried by project members, and share configurations of those adapters that are found to have desirable results.

Our implementation allows for API key management within the project, precipitating security best practices in controlling agents with access to specific adapters. This is accomplished by creating individual API keys for each agent that would access the project's API, so in the case that a key is compromised, it can be easily regenerated, preventing outside actors from accessing the project.

***Secure Computational Network***

As illustrated in Figure 1, our system manages the end-to-end data management, training, and inference for isolated models. This process requires the coordination of data movement, model training, and inference. If our resulting models were intended for public consumption, we could host them on a server exposed to the Internet or utilize a model hosting service such as Hugging Face. If the highest level of security was our goal, we could host individual models on isolated infrastructure, exposing model inference to back-end servers and services. However, if we want to provide large-scale model services that are both flexible and secure, our options are very limited.

The generative nature of LLMs creates the possibility for training data exposure based on input prompts and inference settings. Models and adapters trained with sensitive data must be limited to authorized users. Unfortunately, LLM inference servers are typically used with single models and provide limited, if any, security restrictions, relying on network security to prevent unauthorized access. The lack of granular security is especially problematic in the case of multi-LoRA hosting, where a shared interface is used, and specific adapters are identified per request.

Security implications aside, we still must address the challenge of data and process distribution. We would like to be able to utilize computational resources wherever they are available and without the need for direct network access to training and inference systems. Likewise, we would like to dynamically track the availability, load, and state of known resources and actively discover new resources as they become available.

To satisfy our security, accessibility, and active resource monitoring requirements, we make use of Cresco,[43] an agent-based framework, to establish a computational network overlay between islands of resources. Cresco was developed to support edge-based applications[45] across heterogeneous networks and resources. Key features of the Cresco platform are end-to-end encryption of data, multi-tenant isolation of resources, and the concept of a secure data plane.[46] The Cresco ecosystem is composed of an application description language, software agents, agent plugins, and client libraries. While the details of these implementations are beyond the scope of this targeted paper, we will discuss how key Cresco features are used in this implementation. The overlay network that is established between agent message brokers provides end-to-end secure communication methods through agent-to-agent messages and data streams over the data plane (text and binary data). These resources could be deployed within an institution, a cloud provider, or any location where one or more agents can contact another, forming a mesh of connected resources.

## Results

A key aspect of this system is the ability to provide multi-lora inferencing at time of request, with negligible overhead. Figure 4 shows a sustained load test on a single GPU with request randomly switching between nine adapters across hundreds of prompts with the Llama 2 8B model. As shown in the graph we see an average of 24 responses per second, with an average response time of 1.6 seconds. Scaling inference services across eight independent GPUs on a single node, we can service nearly 200 request per second.

While it is becoming more common for the community to publish adapters, most models are still released with full weights. Thankfully, the MergeKit[47] project provides tools to extract adapters from full fine-tuned weights if the original model is available. For example, Hermes 3[48] is a popular model for function calling, that is based on the Llama 3 model. With MergeKit, the original Llama 3 model, and the Hermes 3 fine-tuned model, we can extract a Hermes 3 adapter. We can then use the Hermes 3 adapter like we would our own custom trained adapters. Given our based model is Llama 3, we can made immediate use of this adapter for function calling on request. In the case of a functional pipeline, the Hermes 3 adapter might be applied during a functional calling stage, while a domain-specific medical adapter might be used to interpret results. As inference server and model technology advances, we hope to transition embedding extraction and multi-model inferencing in the same way, allowing adapters to be used for different functions, even within the same request.

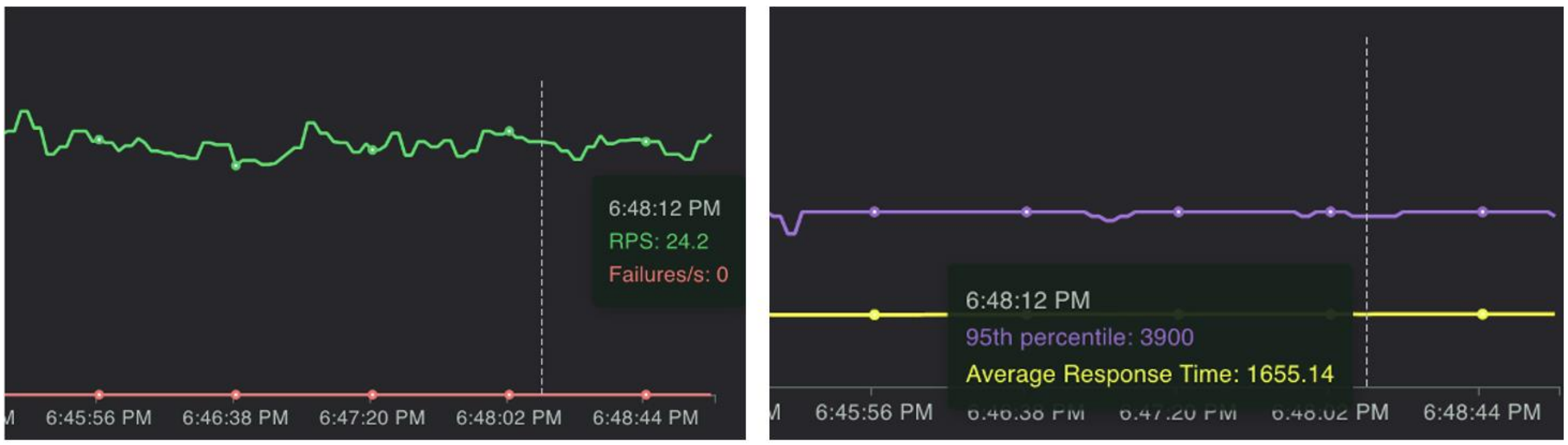

***Figure 4*** *NVIDIA A100 GPU Performance with Llama 3 8B (response time in ms)*

Multi-lora inferencing is not limited to a single adapter. So-called mixture-of-adapters[49] can be used that combine several adapters at time of inferencing with varying weights to influence output. Treating adapter weights like hyperparameters, we used ClearML to optimally determine the mix-of-adapters and associated weights for a set of multiple-choice medical questions. We produced a series of Medical Education Language Transformer (MELT)[50] models demonstrating the benefit of this approach. Using a mix of medical unstructured text, instructional training, and question and answer, and multiple-choice datasets obtained from public sources, we trained four independent adapters. The optimal mix of adapters and weights were evaluated using medqa[51], medmcqa[52], and usmle[53] multiple-choice medical datasets, as shown in Table 1.

| MODEL | BASE MODEL SCORE | MELT MODEL SCORE |
|---|---|---|
| Mixtral-8x7B-Instruct-v0.1 | 65.32% | 68.2% |
| Llama-2-3x70b-chat-hf | 50.37% | 63.6% |
| Mistral-3x7B-Instruct-v0.1 | 44.87% | 53.7% |
| Llama-2-7b-chat-v0.1 | 35.2% | 46.2% |
| TinyLlama-1.1B-Chat-v1.0 | 24.57% | 27.95% |

***Table 1*** *Scores of various base model vs. scores when the MELT adapter was applied*

On average, our mixture-of-adapter optimization approach resulted in a 19% improvement across model types. Additional details related to results for specific models and datasets are provided in the associated model cards. In future efforts we intend on utilizing embeddings to find thematic adapters that are associated with an input question to dynamically optimal inclusion of adapters with associated weights.

## Conclusion

In this paper, we describe a self-service system developed at the University of Kentucky Center for Applied AI to democratize access to large, customized language model resources. Leveraging recent advancements in multi-LoRA inference, allowing hundreds or thousands of adapters to be hosted from shared computational resources, we can efficiently host custom adapters for users across our institution. Through Cresco, a previously developed agent-based system, we can securely bridge islands of resources, including client-facing interfaces, model training, inference, and storage resources. In the described system, we provide secure LLM services through process and data isolation, end-to-end message encryption, and role-based resource authentication. These features facilitate more secure, customizable development and usage of local LLMs, aligning with the stringent privacy regulations that researchers in biomedical informatics must adhere to.

In future efforts, we aim to target the development of tools to assist in dataset and adapter composition. While not yet implemented in available inference servers, the PEFT library allows for the weighted application of multiple adapters within the same inference instance. Conceivably, at execution time, methods could be developed to predict what adapters and specific weights should be used for an incoming request. To accomplish this, we need to rethink how we assemble datasets. For dataset composition, we will continue the development of clustering and classification of individual instruction records across dataset collections. Using existing feature extraction services, semantic themes can be trained into individual adapters of similar thematic clusters. Incoming prompts can be evaluated for adapter similarity based on known adapter datasets, informing execution-time compositions of adapters.

Our intent through this and future related efforts is to provide the research community with self-service access to the latest models and technology as simply and cost-effectively as possible. Building from this and other efforts, we foresee the ability to create networks of participating AI agents, securely providing their own specializations and contributing to broader scientific discovery efforts and provide the biomedical informatics community with new tools with which to promote higher quality healthcare.

Code, examples, and setup instructions can be found in the following repository:
https://github.com/innovationcore/llm_factory_demos

## Acknowledgements

The project described was supported by the University of Kentucky Institute for Biomedical Informatics and the Center for Clinical and Translational Sciences through NIH National Center for Advancing Translational Sciences through grant number UL1TR001998. The content is solely the responsibility of the authors and does not necessarily represent the official views of the NIH.

This material is based upon work supported by the National Science Foundation under Grant No. 2216140.